\documentstyle[11pt,aasms4]{article}
\journalid{}{}
\articleid{}{}
\slugcomment{}

\lefthead{Veilleux \& Bland-Hawthorn}
\righthead{Violent Outflow in Circinus}

\begin{document}

\title{Artillery Shells over Circinus} 

\author{Sylvain Veilleux\altaffilmark{1}
 and Jonathan Bland-Hawthorn\altaffilmark{2}}

\altaffiltext{1}{Department of Astronomy, University of Maryland,
College Park, MD 20742; E-mail: veilleux@astro.umd.edu}

\altaffiltext{2}{Anglo-Australian Observatory, P.O. Box 296, Epping, 
NSW 2121, Australia; E-mail: jbh@aaossz.aao.gov.au}

\begin{abstract}

The recently identified Circinus Galaxy is the nearest ($\sim$ 4 Mpc)
Seyfert 2 galaxy known and we now demonstrate to be one of the best
laboratories for studying the effects of nuclear activity\footnote{In
this paper, we use the terms `nuclear activity' to refer to either
starburst or black-hole driven activity in the nuclei of
galaxies. Similarly, we refer to `active galaxies' as galaxies powered
by star formation (starburst galaxies) or through accretion onto a
massive black hole (active~galactic~nuclei).} on the surrounding
environment.  Here we present new imaging Fabry-Perot observations of
Circinus which confirm the existence of an ionization cone in this
object but also show for the first time a complex of ionized filaments
extending radially from the nucleus out to distances of 1 kpc.  Arcs
suggestive of bow shocks are observed at the terminus of some of these
filamentary structures. Most spectacular of all, one of the structures
appears to be a scaled-up version of a Herbig-Haro jet.  The velocity
field of the filaments confirms that they represent material expelled
from the nucleus (possibly in the form of `bullets') or entrained in a
wide-angle wind roughly aligned with the polar axis of the galaxy.
The motions observed across the ionization cone are highly supersonic,
so high-velocity shocks are likely to contribute to the ionization of
the line emitting gas.  However, it is not clear at present whether
shock ionization dominates over photoionization by the Seyfert 2
nucleus.  Extrapolation of the filaments to smaller radii comes to
within 1\arcsec\ (about 20 pc) of the infrared nucleus, therefore
suggesting a AGN or nuclear starburst origin to these features.  The
complex of radial filaments detected in the Circinus galaxy is unique
among active galaxies. The frequency of such events is unknown since
only a handful of galaxies have been observed at the sensitivity level
of our present observations.  The event in the Circinus galaxy may
represent a relatively common evolutionary phase in the lives of
gas-rich active galaxies during which the dusty cocoon surrounding the
nucleus is expelled by the action of jet or wind phenomena.

\end{abstract}


\section{Introduction}

The Circinus galaxy is a large, isolated, gas-rich spiral seen through
a relatively unobscured ($A_V$ $\approx$ 1.5 mag) optical window near
the Galactic plane ($b$ = --3.8$^\circ$; Lyngo \& Hansson 1972; Mebold
et al. 1976; Freeman et al. 1977).  The early detection of strong
radio emission from the nucleus (Freeman et al. 1977) provided the
first evidence for nuclear activity in this galaxy.  Recent maps
(Harnett et al. 1990; Elmouttie et al. 1995) have resolved spectacular
radio lobes centered on the nucleus, and extending more than
90\arcsec\ ($\sim$ 2 kpc) on either side of the galaxy disk (PA
$\approx$ --60$^\circ$).  Additional evidence of unusual nuclear
activity is suggested by the discovery of an intense 22-GHz H$_2$O
megamaser (Gardner \& Whiteoak 1982; Whiteoak \& Gardner 1986).
Detailed spectroscopic studies at optical and infrared wavelengths
have since revealed the presence of a Seyfert 2 nucleus surrounded by
an extended (200-pc radius) circumnuclear starburst (Oliva et
al. 1994, 1995; Ghosh 1992; Marconi et al. 1995). A one-sided ionization
cone and gas motions indicative of a large central mass concentration
have recently been discovered in the Circinus galaxy (Marconi et
al. 1995; Greenhill et al. 1997), making it the closest galaxy where the engine
responsible for the nuclear activity can convincingly be attributed to
a supermassive black hole surrounded by a thick obscuring screen. In
this {\it Letter}, we describe new {\it TAURUS-2} and long-slit
spectroscopic observations which reveal a complex of radial filaments
emanating from the nucleus of this galaxy. We propose two possible
scenarios to explain these features and briefly discuss the
implications of our new results.

\section{Observations}

The Circinus galaxy was observed for 165 minutes on the night of
February 21 1995 using the {\it TAURUS-2}~imaging Fabry-Perot
interferometer at the AAT 3.9-m telescope. This instrument was used in
the Angstrom imaging mode to maximize sensitivity to faint line
emission (Bland-Hawthorn et al. 1997). A 40$\mu$m gap etalon was used
out of band to produce deep, low-resolution ($\sim$ 350 km s$^{-1}$
sampled at every $\sim$ 30 km s$^{-1}$) [O~III] $\lambda$5007 spectra
at each point across this galaxy.  Each square pixel subtended 0\farcs
315 on the sky; the atmospheric seeing at FWHM averaged approximately
4 times this value.  These data comprise kinematic and photometric
information for the [O~III] line at $\sim$ 10,000 positions over the
central 2\arcmin\ of the Circinus galaxy.  At a later stage, Circinus
was observed at several other etalon spacings to derive the
distribution of H$\alpha$ emission and construct line profiles of
[S~II] $\lambda\lambda$6716, 6731 and [O~III] $\lambda$5007 at 1
\AA~resolution. Narrow-band filters were also used at different tilt
angles to isolate the lines of He~II $\lambda$4686, [S~II]
$\lambda\lambda$6717,6731 and H$\alpha$, and in order to subtract
neighbouring continuum emission.  The H$\alpha$ flux map is discussed
below, but the rest of these data are to be presented in a more
detailed study (in progress).

In addition, S.L. Lumsden kindly obtained long-slit observations for us at two
positions in Circinus. The RGO spectrograph on the AAT was used at the
Cassegrain f/8 focus with the 25cm camera and the 270R grating. At a
plate scale 0\farcs 77 pix$^{-1}$, this set up gave a resolution of
3.4\AA\ FWHM at H$\alpha$ and a wavelength coverage of
4770 -- 8260 \AA. The 2\arcsec\ slit was somewhat
oversized compared to the 1\farcs 1 FWHM seeing. Both slits were
aligned at a position angle of 270$^\circ$, one passing through the
nucleus, the other offset 6\farcs 0 N of the nucleus. The exposure
time for both positions and an offset sky position was 600 sec.

\section{Results}

Figures 1 and 2 show the [O~III] and H$\alpha$ line flux images
obtained by integrating the line profiles in the Fabry-Pero data. Only
the blueshifted (between --150 and 0 km s$^{-1}$) H$\alpha$ emission
is presented in these figures to minimize contamination from the
(redshifted) circumnuclear emission south-west of the nucleus. These
data confirm the existence of a bright ionization cone in Circinus
(Marconi et al. 1995), an effect also apparent in our He~II and [S~II]
narrow-band images.  But the [O~III] data also reveal a number of
fainter filaments beyond this cone which cannot be explained by simple
illumination effects of an homogeneous environment by an anisotropic
(biconical) source of radiation in the nucleus.

The most striking [O~III] feature extends along position angle $\sim$
--50$^\circ$ spanning a distance of $\sim$ 25 -- 45\arcsec\ (500 -- 900
pc) from the nucleus. The lateral extent of this filament is near the
limit of our resolution ($\sim$ 1\farcs 5 after smoothing). This
narrow feature is also visible at H$\alpha$ but with a lower
contrast. The gas at this location is highly ionized with a [O~III]
$\lambda$5007/H$\alpha$ flux ratio typically larger than
unity. Extrapolation of this filament to smaller radii comes to within
1\arcsec\ (20 pc) of the infrared nucleus (Marconi et al. 1995),
suggesting a nuclear (AGN or compact starburst) origin to this
feature. A second [O~III] filament is also detected, emerging from the
nucleus along PA $\approx$ --120$^\circ$ out to a maximum radius of
$\sim$ 35\arcsec\ (700 pc).  These radial filaments resemble the
optical counterparts of radio jets in more powerful active galaxies
(e.g., Sparks, W. B., Biretta, J. A., \& Macchetto, F. 1994).  

The most spectacular feature in the H$\alpha$ data is the hook-shaped
filament which extends to 40\arcsec\ (800 pc) west of the nucleus
(Fig. 2).  Such features are commonly observed in Herbig-Haro (HH)
objects (e.g., HH 47; Hartigan et al. 1993) although have never been
seen on galactic scales.  The western `hook' is far more elongated
than the wind-blown bubbles in M82 (Bland \& Tully 1988) and NGC 3079
(Veilleux et al. 1994).  Additional morphological evidence for outflow
exists in the northern portion of our data (Fig. 1).  The [O~III]
emission along PA $\approx$ --20$^\circ$ forms a broad filamentary
`finger' or jet that points back to the nucleus. A knot is present at
the tip of this `finger', 25\arcsec\ from the nucleus.  Bright H$\alpha$
emission is also visible near this position, the southern portion of
which forms a wide ($\sim$ 8\arcsec\ ) arc resembling a bow shock. The
arc is pointing in the downstream direction consistent with being
produced by a collimated jet.

The kinematics derived from the [O~III] Fabry-Perot data (Fig. 3)
and long-slit spectra (Figs. 4 \& 5) bring credence to this nuclear outflow
scenario.  Non-gravitational motions are observed throughout the
[O~III] cone, superposed on a large-scale velocity gradient caused by
galactic rotation along the major axis of the galaxy (PA$_{\rm maj}$
$\approx$ 30$^\circ$; Freeman et al. 1977).  An unusually large
velocity gradient of 4 km s$^{-1}$ pc$^{-1}$ is seen near the position
of the bright knot $\sim$ 12\arcsec\ from the nucleus along PA
$\approx$ --30$^\circ$ (Fig. 3).  The side of the knot facing the
nucleus presents velocities that are nearly 250 km s$^{-1}$ lower than
gas only 3\arcsec\ north of that position.  The emission profiles near
that knot are broad ($\sim$ 250 km s$^{-1}$) and perhaps
complex.

The material in the brighter portions of the NW and SW filaments does
not seem to take part in the galactic rotation (Fig. 3).  The
velocities in the NW [O~III] filament appear systematically
blueshifted by $\sim$ 0 -- 100 km s$^{-1}$ with respect to the
systemic velocity (439 km s$^{-1}$; Freeman et al. 1977), while the
velocities of the gas in the SW filament are roughly
systemic within the errors of the measurements.
Non-gravitational motions are also detected along the
western `hook' feature (Fig. 4). A velocity gradient of 80 km s$^{-1}$
over 8\arcsec\ (0.5 km s$^{-1}$ pc$^{-1}$) is visible near the
location of knot 2 (Fig. 4a). 
Perhaps the strongest evidence for shocks in our data is seen in knot
4. There, the strong H$\alpha$ line is blueshifted by 180 km
s$^{-1}$ with respect to the [NII] doublet (Figs. 4b \& 5).  Velocity
shifts between different emission lines are frequently observed in HH
objects and reflect spatially distinct line-emitting regions (e.g.,
leading edge versus cooling tail; Morse et al. 1994 and references
therein).  A more detailed analysis of the nuclear long-slit spectra
also suggests the presence of broad (FWZI $\approx$ 800 km s$^{-1}$)
blueshifted wings in the emission-line profiles produced by knots 2
and 3, but this result needs to be confirmed with spectra of higher
signal-to-noise ratio and velocity resolution.

The current radio data also support the existence of a
wide-angle outflow in the Circinus galaxy.  The NW feature appears to
have a radio counterpart at both 13cm and 20cm (the NW `plume' in
Figs. 2$-$4 of Elmouttie et al 1995).  This appears as a continuum
ridge which runs SE$-$NW through the radio map out to 90\arcsec\ in
radius within the bisymmetric lobes.  Radio jets in active galaxies
are commonly observed to have associated optical emission, but
normally confined to an outer surface at the terminus of the shock
(Cecil, Bland, \& Tully 1990).  The radio data do not have sufficient
resolution to argue whether the optical emission fills the NW `plume'
or is confined to the shock front.  The 13cm
radio map also shows a narrow feature which extends due west from the
nucleus. At higher resolution, we anticipate that this feature is
associated with another jet $-$ distinct from the NW jet $-$
responsible for the western optical `hook'.  Any possible radio
counterparts to the SW filament and the northern `finger' are being
masked by strong emission from the galactic disk.

The motions observed across the ionization cone are highly supersonic,
so high-velocity ($V_s$~$\ga$~100~km~s$^{-1}$) shocks are likely to
contribute to the ionization of the line emitting gas.  Table 1
summarizes the line ratios measured from our long-slit spectra in
knots 1 -- 4 of the western `hook' and in the nucleus. Large
variations are sometimes observed within a single knot. In knot 2, for
instance, [O~I]/H$\alpha$ falls with radius as [O~III]/H$\beta$
increases. This gradient argues against photoionization (of
ionization-bounded clouds) by the nucleus, unless the ionization
parameter is rising due to a rapidly falling density profile. The
relatively constant density-sensitive [S~II]
$\lambda$6717/$\lambda$6731 ratio measured at these locations
(especially at the two eastern positions in knot 2; cf. Table 1,
off-nuclear spectrum) rules out this possibility.  The very different
excitation properties of knots 3 and 4 (cf. Table 1, nuclear spectrum;
Fig. 5) are also difficult to explain in the pure nuclear
photoionization scenario, although here large variations of the
ionization parameter associated with density variations cannot
formally be excluded because the [S~II] $\lambda$6716/$\lambda$6731
ratio is in the low-density limit and therefore cannot be used as a
density indicator.  The enhanced [N~II]/H$\alpha$, [S~II]/H$\alpha$, 
and [O~III]/H$\beta$ ratios in knots 1 and 2 fall near the range
produced by the high velocity photoionizing radiative shocks of Dopita
\& Sutherland (1995).  However, [O~I]/H$\alpha$ is considerably {\it weaker}
than predicted by the models. Moreover, all of these ratios require
large gas velocities (V$_s$ $\ga$ 500 km s$^{-1}$) and therefore imply
large projection effects to explain the relatively small apparent
velocities in our data.  Hybrid models involving fast shocks and
power-law photoionization by the Seyfert 2 nucleus (e.g.,
Viegas-Aldrovandi \& Contini 1989 and references therein) alleviate
these problems. Photoionization by the active nucleus of a mixture of
ionization and matter-bounded clouds whose relative proportions vary
with position in the galaxy may also provide another explanation
for the abrupt changes of excitation in the filaments (e.g., Binette,
Wilson, \& Storchi-Bergmann 1996).

\section{Origin of the Complex of Radial Filaments}

Jets in active galactic nuclei are attributed to gas centrifugally
accelerated along magnetic field lines tied at one end to the
accretion disks (Lynden-Bell 1996).  The existence of several jets in
a single galaxy is difficult to explain in these models unless the
active galaxy is host to two or more black holes each possessing its
own accretion disk and radio jet.  A more plausible explanation for
the optical filaments and the linear structures observed in the radio
map is that they arise from individual mass structures ejected in a
wide opening angle possibly from an explosive nuclear event.  The
apparent extent and velocities of the outflowing gas in the Circinus
galaxy suggest that the purported explosive event took place a few
million years ago.  It is not clear what triggered this event since
the Circinus galaxy shows no sign of recent galactic interaction
(Freeman et al. 1977).  The ejecta from this event have been funneled
into a fan with opening angle of $\sim$ 100$^\circ$ and symmetry axis
along PA $\approx$ --65$^\circ$, i.e.  along the direction of the NW
radio lobe (Elmouttie et al. 1995).  The lack of any optical
counterpart to the SE radio lobe argues that the outflow on that side
is hidden from view by the inclined ($\sim$ 65$^\circ$; Freeman et
al. 1977) galaxy disk. The symmetry axis of the conical outflow is
roughly aligned with the minor axis of the galaxy, and therefore
suggests that our line of sight lies only $\sim$ 15$^\circ$ outside
the cone defined by the outflow.

Hydrodynamical instabilities in the dense shell swept up by an
expanding plasmon (Pedlar, Dyson, \& Unger 1985; Taylor et al. 1989)
or a time-variable wind (Garcia-Segura, MacLow, \& Langer 1996; Stone,
Xu, \& Mundy 1995) offer an alternative explanation for the optical
morphology of the Circinus galaxy.  Dynamical models (MacLow, McCray,
\& Norman 1989; Tomisaka \& Ikeuchi 1988; Suchkov et al. 1994) of
wind-blown bubbles have had success reproducing the range of phenomena
observed in radio-lobe spirals (Wehrle \& Morris 1988; Wrobel 1994;
Veilleux et al. 1994).  The outflow event in the Circinus galaxy would
fundamentally be driven by this same mechanism, but would correspond
to a later stage of evolution when the initial wind-blown shell has
broken out and left dense clouds that have since been accelerated and
stretched radially by the galactic wind (free wind phase).  The impact
of a high-speed supersonic wind on dense clouds has been discussed in
several contexts including shock-induced star formation (Woodward
1976), supernova blast waves (McKee et al. 1987), and broad-absorption
line quasars (Schiano, Christiansen, \& Knerr 1995).  Numerical
simulations indicate that the initial encounter of the clouds with the
wind medium drives a strong shock that may have devastating effects on
the cloud structure (Woodward 1976; McKee et al. 1987). Once in ram
pressure equilibrium with the wind, however, these clouds may be
accelerated up to a significant fraction of the wind velocity before
Rayleigh-Taylor and Kelvin-Helmholtz instabilities tear them
apart (Schiano et al. 1995).

A clue to the origin of the outflow in the Circinus galaxy is provided
by the energetics of this event.  The mass taking part in this outflow
can be estimated from the [O~III] $\lambda$5007 flux.  We parametrize
the mass in terms of the density, which is poorly constrained, and use
an electron temperature of 10$^4$ K. The integrated observed [O~III]
intensity of $\sim$ 5 $\times$ 10$^{-13}$ erg s$^{-1}$ cm$^{-2}$
implies a mass of a few times 10$^4$ X$^{-1}$ N$_{e,2}^{-1}$ M$_\odot$
where X is the fraction ($<$ 1) of oxygen which is doubly ionized and
N$_{e,2}$ is the electron density in units of 10$^2$ cm$^{-3}$. In
this calculation, we used A$_v$ = 1.5 mag, a solar mass fraction for
oxygen, and a five-level atom approximation to estimate the emission
coefficient (McCall 1984).  Taking 200 km s$^{-1}$ as representative
for the deprojected gas velocities (a uncertain number; cf. last
paragraph of \S 3), the kinetic energy involved in the NW outflow is a
few times 10$^{52}$ N$_{e,2}^{-1}$ ergs. A value closer to $\sim$
10$^{53}$ N$_{e,2}^{-1}$ ergs is probably more representative of the
total energy involved in the outflow event since the radio data
(Harnett et al. 1990; Elmouttie et al. 1995) suggest that a similar
mass outflow is also taking place on the SE side of the nucleus.  The
energetics of the optical outflow in the Circinus galaxy therefore
appear to be relatively modest (equivalent to about 100 supernova
explosions if N$_e$ $\approx$ 100 cm$^{-3}$) and appear to lie at the
low energy end of the distribution for wide-angle events observed in
nearby galaxies (Cecil et al. 1990; Bland \& Tully 1988; Heckman et
al. 1990; Veilleux et al. 1994).  This outflow can easily be powered
by the AGN or by a compact nuclear starburst. The nuclear starburst
detected in Circinus (Oliva et al. 1995) appears somewhat older than
the present outflow event, however.

\section{Summary and Implications}

Deep imaging Fabry-Perot data reveal a complex of radial line-emitting
filaments in the Circinus galaxy, the closest Seyfert 2 galaxy
known. The kinematics of the gas producing these features suggest the
ejection of material over a wide opening angle or inhomogeneities in a
wide-angle outflow.  The proximity of the Circinus galaxy makes it a
unique laboratory to study with unprecedented resolution the impact of
nuclear winds, supersonic ejecta, and jets on the interstellar medium
of galaxies.  The apparent rarity of radial filaments and bow shocks
in active galaxies may reflect the ephemeral nature of
this phenomenon or the difficulty in accelerating dense gas clouds
coherently.  However, few galaxies have been observed with the
sensitivity of our present observations. The discovery of these
features in the Circinus galaxy, a spiral galaxy with an abnormal
richness of gas (Freeman et al. 1977), brings up the possibility that
we may be witnessing a common evolutionary phase in the lives of
gas-rich active galaxies. Observations of other active galaxies at
similar sensitivity will help establish the frequency and duration of
this phenomenon.

\clearpage

\acknowledgments

We acknowledge useful conversations with Roger Blandford, Pat
Hartigan, and Jim Stone on the physics of outflows.  We would
particularly like to thank S. L. Lumsden for obtaining the long-slit
spectra presented in this paper, Keith Jones for helpful comments on
the radio data, Brent Tully for the loan of the HIFI etalon, and the
referee, E. Oliva, for several suggestions which improved this paper.
Parts of this work have been supported by NASA through grant number
HF-1039.01-92A awarded by the Space Telescope Science Institute which
is operated by the AURA, Inc. for NASA under contract No. NAS5--26555
(SV).

\clearpage

\begin{figure}
\epsscale{0.5}
\caption{ Line flux images of the Circinus galaxy: $a$, [O~III]
$\lambda$5007 and $b$, blueshifted (between --150 and 0 km s$^{-1}$) H$\alpha$.
North is at the top and west to the right.  The position of the
infrared nucleus (Marconi et al. 1995) is indicated in each image by a
cross. The spatial scale, indicated by a horizontal bar at the bottom
of the [O~III] image, is the same for each image and corresponds to
$\sim$ 25\arcsec\, or 500 pc for the adopted distance of the Circinus
galaxy of 4 Mpc.  The minor axis of the galaxy runs along PA $\approx$
60$^\circ$ (measured from north to east). The faintest features in the
[O~III] (H$\alpha$) image have a surface brightness of $\sim$ 4 (20)
$\times$ 10$^{-17}$ erg s$^{-1}$ cm$^{-2}$ arcsec$^{-2}$.  To suppress
the wide dynamic range, the intensity greyscale wraps around and
becomes logarithmic at high intensity levels. The radial features
along the north (labelled N), north-west, west (labelled W), and
south-west axes suggest the ejection of material (possibly in the form
of `bullets') over a wide opening angle or inhomogeneities in a
wide-angle outflow.  The inclined ($\sim$ 65$^\circ$; Freeman et
al. 1977) galaxy disk hides the south-east portion of this outflow.  }
\end{figure}

\begin{figure}
\epsscale{0.75}
\caption{ Line flux images of the western hook-shaped filament: $a$,
[O~III] $\lambda$5007 and $b$, blueshifted (between --150 and 0 km s$^{-1}$)
H$\alpha$. The position of the infrared nucleus (Marconi et al. 1995)
is indicated in each image by a cross.  The orientation is the same as
in Fig. 1, but the horizontal bar at the bottom of the [O~III] image
now corresponds to $\sim$ 250 pc.  The faintest features in the
[O~III] (H$\alpha$) image have a surface brightness of $\sim$ 1 (4)
$\times$ 10$^{-16}$ erg s$^{-1}$ cm$^{-2}$ arcsec$^{-2}$.  Once again,
the intensity greyscale wraps around and becomes logarithmic at high
intensity levels. The hook-shaped filament shares a strong resemblance
with HH objects produced by young stars, suggesting a bow-shock origin
for this feature.  The bow-shock terminus (labelled 4 in the figure)
is only detected in H$\alpha$, but a rough correspondence is observed
between the [O~III] and H$\alpha$ knots (labelled 1 -- 3) which
delineate the northern edge of this structure. }
\end{figure}

\begin{figure}
\epsscale{0.75}
\caption{ Velocity field derived from the [OIII] data cube.  The
velocities range from 170 km s$^{-1}$ (white) to 650 km s$^{-1}$
(black). The uncertainties
range from about 50 km s$^{-1}$ in the bright line-emitting regions to 
100 km s$^{-1}$ or more in the fainter areas. The
position of the infrared nucleus (Marconi et al. 1995) is indicated by
a cross.  The orientation is the same as in Fig. 1, but the horizontal
bar at the bottom of the image now corresponds to $\sim$ 250 pc.  The large-scale velocity gradient along PA $\approx$
30$^\circ$ is due to galactic rotation. Superposed on this gradient
are non-gravitational motions observed
at several locations, including the NW, W, and SW filaments 
visible in Figure 1, and a compact region $\sim$ 12\arcsec\ from the
nucleus along PA $\approx$ --30$^\circ$ coinciding with very strong
[O~III] emission. }
\end{figure}

\begin{figure}
\epsscale{0.75}
\caption{ Sky-subtracted long-slit spectra. (a) 6\farcs 0 north of
nucleus, PA = 270$^\circ$, (b) through nucleus, PA = 270$^\circ$. The
vertical bar on the right of each panel represents $\sim$ 250 pc. The
off-nuclear spectrum intersects knots 1 and 2 of Fig. 2, while the
nuclear spectrum intersects knots 3 and 4 but only the southern
portions of knots 1 and 2.  Note the velocity gradient in knot 2 of
the upper panel and the strong blueshifted H$\alpha$ emission in knot
4.}
\end{figure}

\begin{figure}
\epsscale{0.75}
\caption{ 
Binned spectra at four positions along the slit through the nucleus
corresponding to knots 1$-$4. Each spectrum has been offset by 400
counts. Knots 1 and 2 are binned along the slit over 6\arcsec; knots 3
and 4 are binned over 4\arcsec. Note the greatly enhanced [N~II]/H$\alpha$
ratio along the jet, except on the bow shock at knot 4. Furthermore, 
the H$\alpha$ line at knot 4 is blueshifted by 180 km s$^{-1}$ with respect to
the [N~II] lines. A faint blue wing can be seen on many of the emission 
lines.}
\end{figure}

\clearpage

\setcounter{figure}{0}
\begin{figure}
\epsscale{0.5}
\plotone{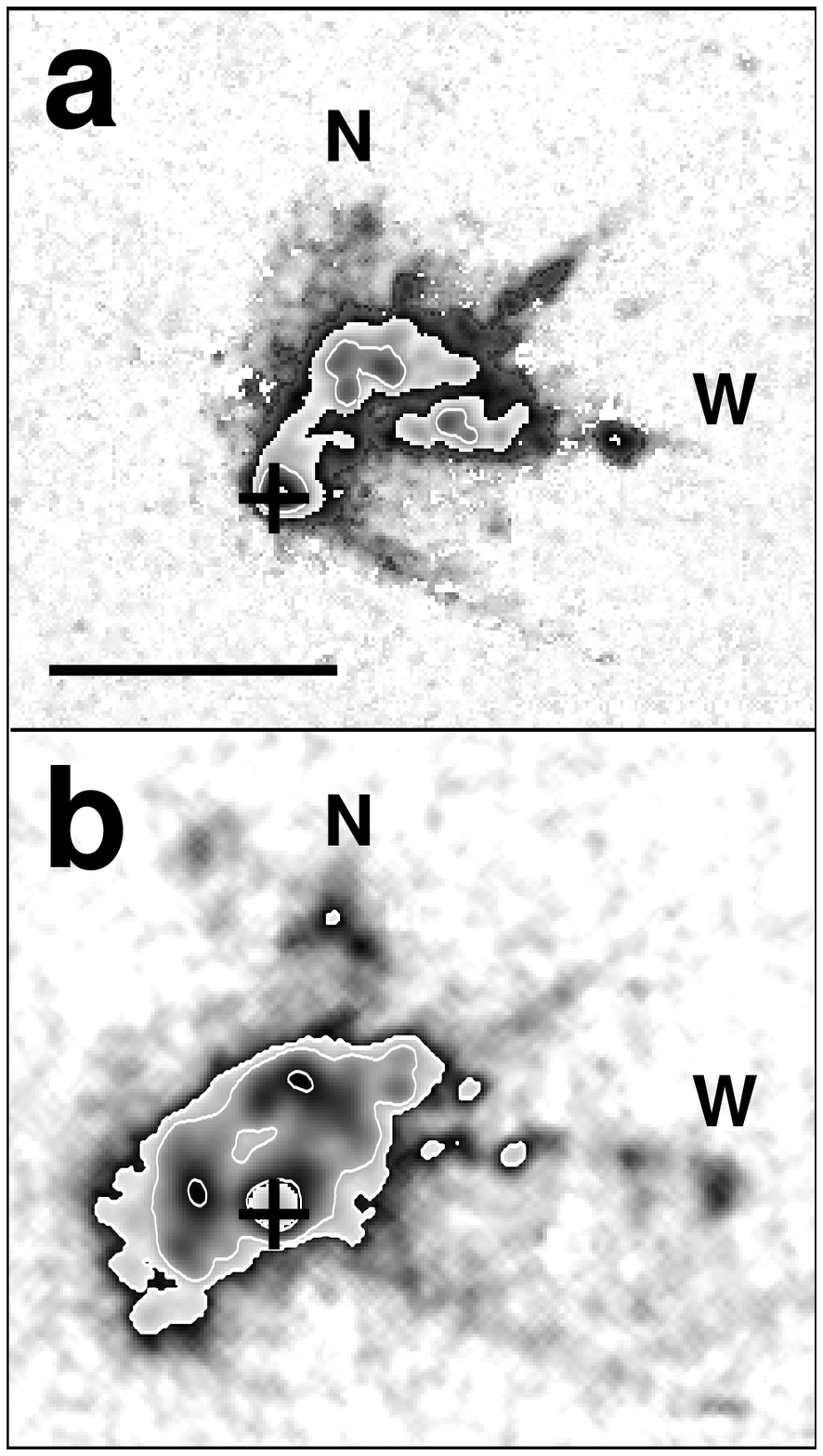}
\caption{}
\end{figure}

\begin{figure}
\epsscale{0.75}
\plotone{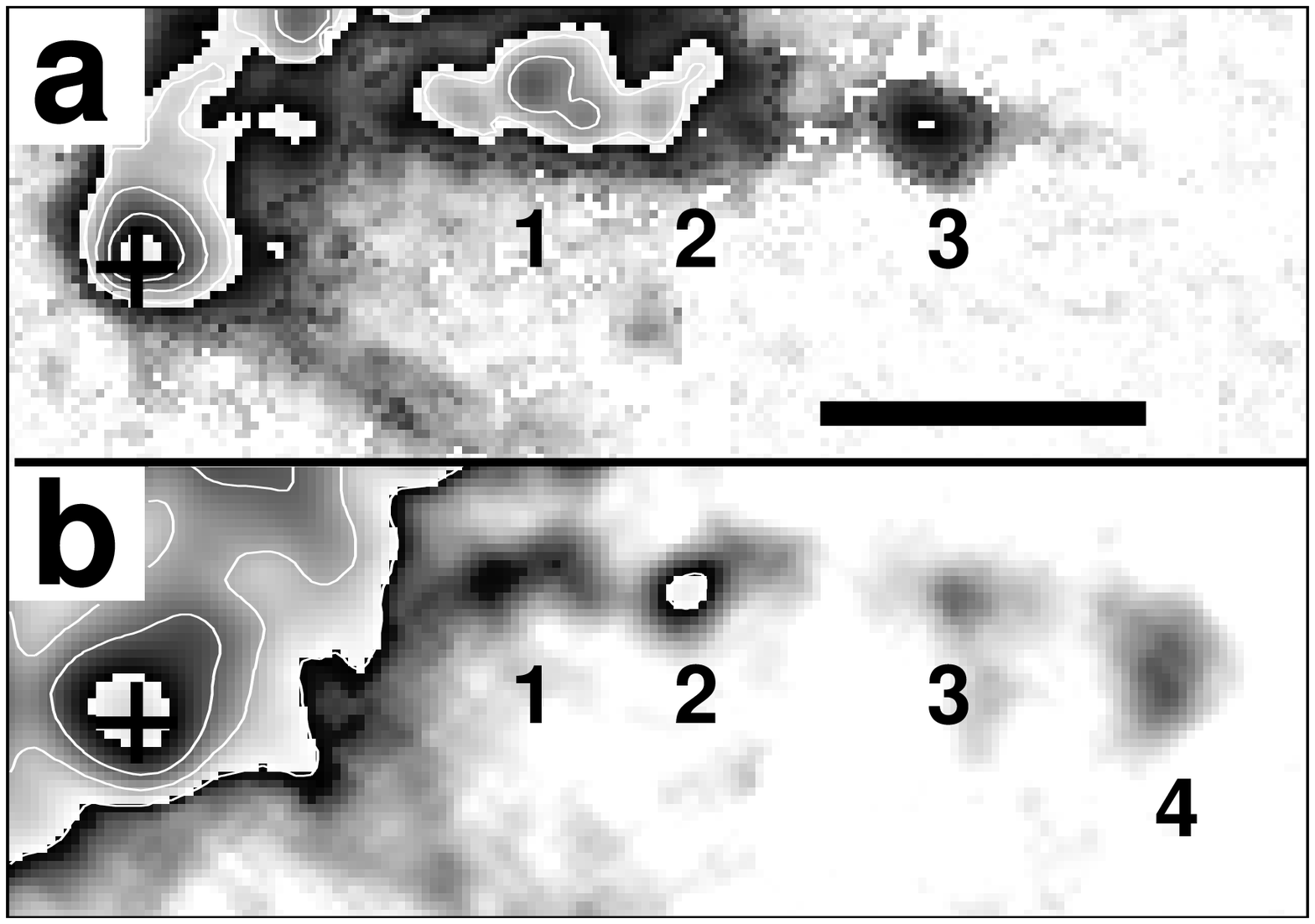}
\caption{}
\end{figure}

\begin{figure}
\epsscale{0.75}
\plotone{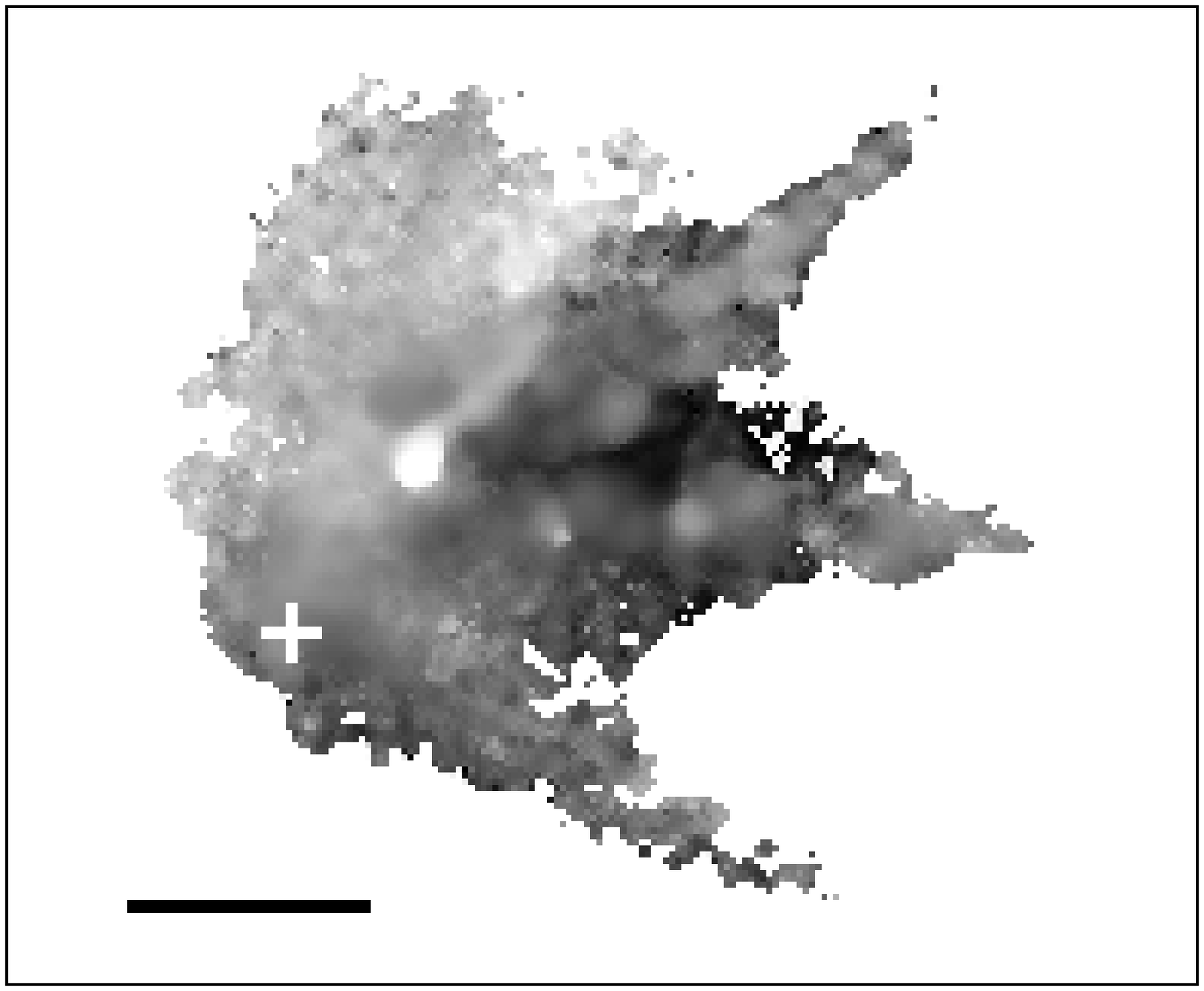}
\caption{}
\end{figure}

\clearpage

\begin{figure}
\epsscale{0.75}
\plotone{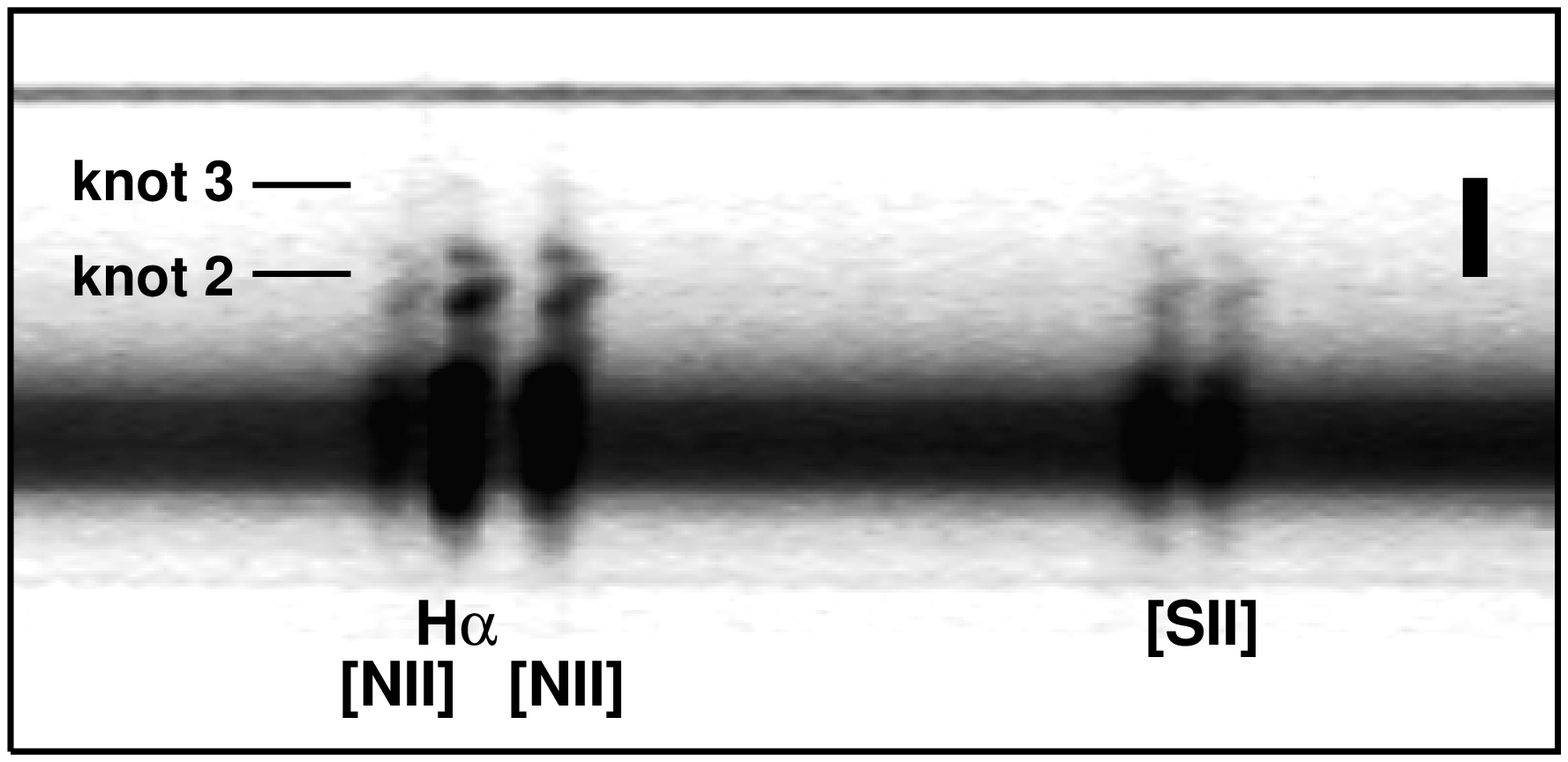}
\caption{a.}
\end{figure}

\setcounter{figure}{3}
\begin{figure}
\epsscale{0.75}
\plotone{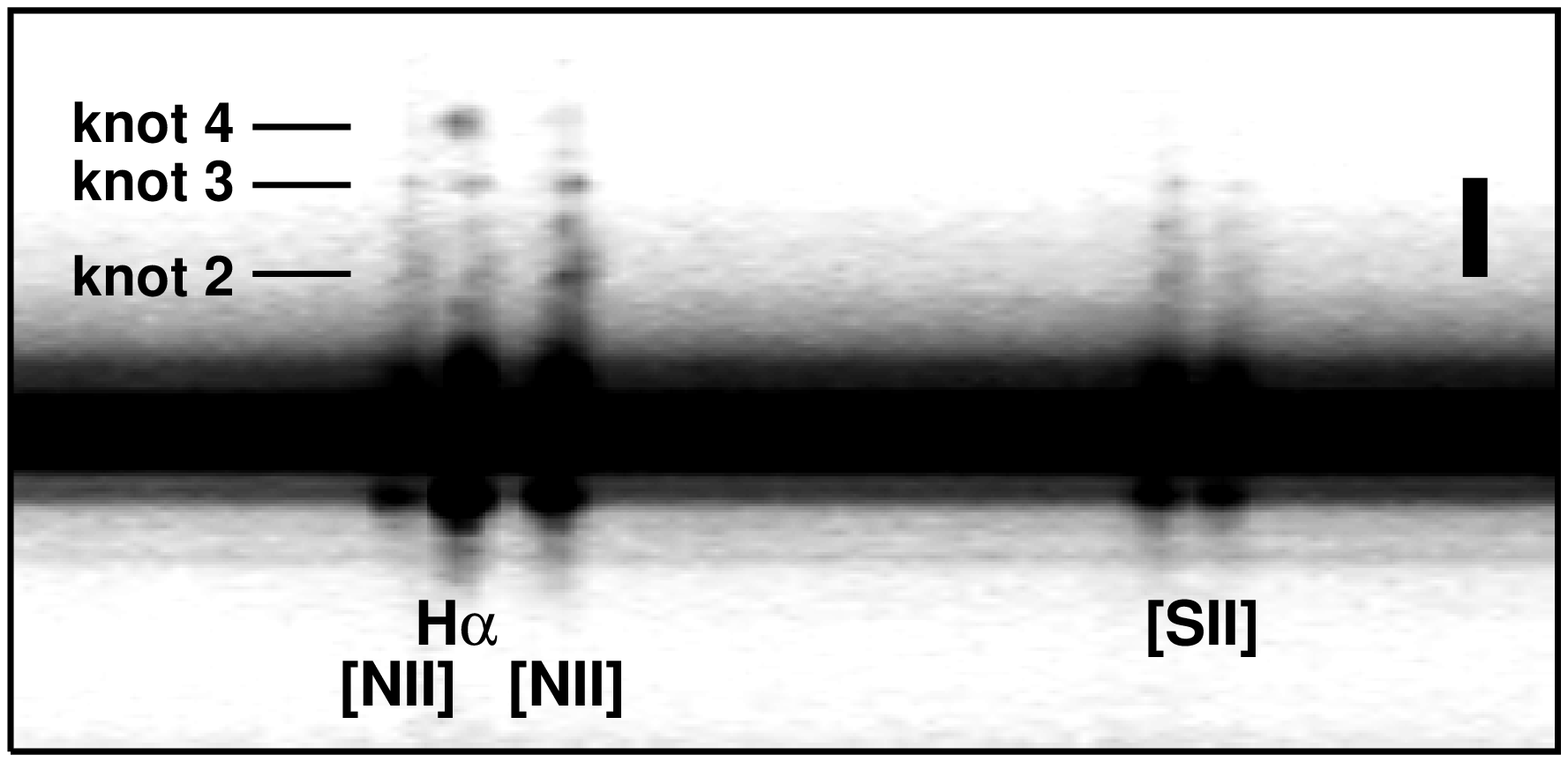}
\caption{b.}
\end{figure}

\begin{figure}
\epsscale{0.75}
\plotone{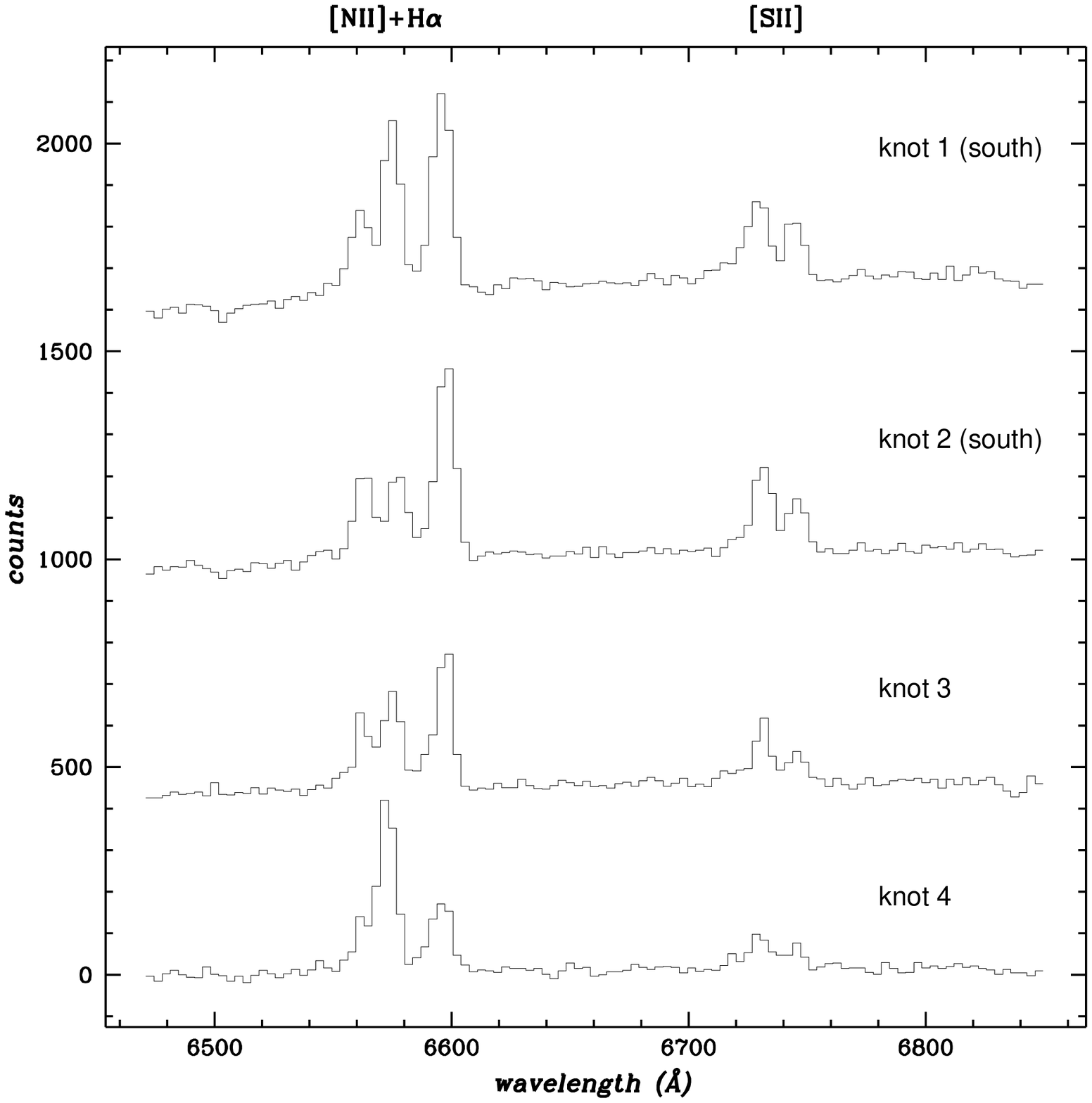}
\caption{}
\end{figure}

\clearpage

\tiny

\begin{table*}
\tablenum{1}
\caption{Emission-Line Ratios derived from Long-Slit Spectra$^{(1)}$.}
\label{tbl-1}

\begin{center}
\begin{tabular}{llccccc}
\tableline
\tableline
 & & & & & & \\
Feature & Region & [O~III]/H$\beta$ & [O~I]/H$\alpha$ & [N~II]/H$\alpha$ & [S~II]/H$\alpha$ & [S~II] 6717/6731 \\
 & & & & & \\
(1) & (2) & (3) & (4) & (5) & (6) & (7)\\
\tableline
 \noalign{\vskip 7.5pt}  
 \multicolumn{7}{c}{Off-Nuclear Spectrum: 6\farcs 0 north of nucleus, PA = 270$^\circ$} \\
 \noalign{\vskip 7.5pt}  
Disk  &  $-$10.0 -- 10.0& 8.94& 0.23& 0.84& 0.60& 1.38 \cr
Knot 1 &  10.8 -- 13.9& 10.0& 0.32& 0.99& 0.73& 1.78 \cr
Knot 2 &  14.6 -- 16.2& 13.9& 0.23& 0.89& 0.58& 1.22\cr
Knot 2 &  16.9 -- 20.0& 16.9& 0.13& 0.75& 0.45& 1.27\cr
Knot 2 &  20.8 -- 22.3& 17.2& 0.13& 0.97& 0.59& 1.54 \cr
 \noalign{\vskip 7.5pt}  
 \multicolumn{7}{c}{Nuclear Spectrum: through nucleus, PA = 270$^\circ$}\\
 \noalign{\vskip 7.5pt}  
Nucleus &$-$3.9 -- 3.9& 13.9& 0.12& 1.12& 0.28& 1.20 \cr
Knot 1 &   11.6 -- 17.7& ...$^{(2)}$& 0.13& 1.41& 0.93:& 2.0: \cr
Knot 2 &   18.5 -- 23.9& ...$^{(2)}$& 0.8:& 2.00& 1.42& 1.85\cr
Knot 3 &   30.8 -- 33.9& ...$^{(2)}$&  ...$^{(3)}$& 1.10& 0.67& 1.80\cr
Knot 4 &   37.7 -- 42.4& ...$^{(2)}$&  ...$^{(3)}$& 0.28& 0.35:& 2.0:\cr
 \noalign{\vskip 7.5pt}  
\tableline
\end{tabular}
\end{center}

\tablenotetext{\bullet}{Meaning of columns:}

\tablenotetext{}{Column (1) -- Name of the emission-line feature following 
the nomenclature of Figure 2. }

\tablenotetext{}{Column (2) -- Region from the long-slit spectrum
which was used to calculate the line ratios. The numbers listed in this
column represent distances in arcseconds measured west of the nucleus. }

\tablenotetext{}{Column (3) -- Flux ratio F([O~III] $\lambda$5007)/F(H$\beta$).}

\tablenotetext{}{Column (4) -- Flux ratio F([O~I] $\lambda$6300)/F(H$\alpha$).}

\tablenotetext{}{Column (5) -- Flux ratio F([N~II] $\lambda$6583)/F(H$\alpha$).}

\tablenotetext{}{Column (6) -- Flux ratio F([S~II] $\lambda\lambda$6717, 6731)/F(H$\alpha$).}

\tablenotetext{}{Column (7) -- Flux ratio F([S~II] $\lambda$6717)/F([S~II] $\lambda$6731).}

\tablenotetext{}{}

\tablenotetext{^{(1)}}{Uncertainties on the line ratios range from 10 -- 15\%
for the stronger lines to 20\% for ratios involving [S~II], and is 
sometimes as high as 40\% for [O~I]/H$\alpha$.}

\tablenotetext{^{(2)}}{H$\beta$ undetected except perhaps for faint
broad emission from knots 3 and 4.}

\tablenotetext{^{(3)}}{[O~I] $\lambda$6300 undetected.}

\end{table*}

\end{document}